\documentclass[letter]{aa} 

\usepackage{graphicx, comment}
\usepackage{commath}
\usepackage{txfonts}
\usepackage{subfig}
\usepackage{upgreek}
\usepackage{threeparttable}
\usepackage{booktabs}      

\usepackage[colorlinks=true]{hyperref}

\hypersetup{
     colorlinks   = true,
     citecolor    = blue
}

\usepackage{orcidlink}

\newcommand{\OJ}{OJ\,248}
\newcommand{\beam}{0.24\,\textrm{mas}}
\newcommand{\rms}{0.71\,\textrm{mJy/beam}}
\newcommand{\vel}{0.17\pm0.06\,\textrm{mas/yr}}

\begin{document} 
\renewcommand{\arraystretch}{1.5}

    \title{A shocking outcome: Jet dynamics and polarimetric signatures of the multi-band flare in blazar OJ\,248}

   \author{
   G.~F. Paraschos\inst{1} \orcidlink{0000-0001-6757-3098}
          }

   \authorrunning{G.~F. Paraschos}
   \institute{
              $^{1}$Max-Planck-Institut f\"ur Radioastronomie, Auf dem H\"ugel 69, D-53121 Bonn, Germany\\ 
              $^{}$\ \email{gfparaschos@mpifr-bonn.mpg.de}\\
             }

   \date{Received -; accepted -}

 \abstract
   {
   The connection between $\gamma$-ray flares and blazars is a topic of active research, with few sources exhibiting distinct enough such outbursts to be able to conclusively connect them to features in their jet morphology.
   Here we present an investigation of the sole $\gamma$-ray flare of the blazar \OJ\ thus far, in association with its jet structure, as revealed by very long baseline interferometry (VLBI).
   We find that throughout the course of the $\gamma$-ray flare, the fractional linear polarisation increases in the jet of \OJ, and the VLBI electric vector position angles (EVPAs) turn perpendicular to the bulk jet flow.
   We interpret this behaviour as a moving shock, travelling through a recollimation shock and upscattering photons via the inverse Compton scattering process, producing a $\gamma$-ray flare; we discuss possible mechanisms.  
   Our hypothesised shock-shock interaction scenario is a viable mechanism to induce such EVPA rotations in both optical and radio bands.
   }


   \maketitle

\section{Introduction}

The flat spectrum radio quasar (FSRQ) \OJ\ (0827+243), located at \(z=0.939\) \citep{Hewett10}, is noted to be a bright, \(\gamma\)-ray loud blazar \citep{Abdo10}.
Its powerful radio jet, propelled by a central supermassive black hole \citep[\(M_\textrm{dyn}\sim8\times10^8\,M_\odot\)][]{Zhang24}, has been resolved at multiple scales with very long baseline interferometry (VLBI).
At parsec-scales the \OJ\ jet exhibits superluminal motion \citep{Jorstad01, Piner06}, while at larger scales \citep{Price93} its bent morphology coincides with X-ray emission detected with the \textit{Chandra X-ray Observatory} \citep{Jorstad04b, Jorstad06}.

Variability has also been observed in \OJ, both in total intensity optical and near infrared light curves \citep[in terms of flaring activity, e.g.][]{Villata97, Raiteri98, Enya02, Marchesini16}, as well as at VLBI scales \citep[in terms of morphological changes, e.g.][]{Marscher83b}.
Such VLBI-detected structural changes in the jet morphology are thought to be caused by the dominating magnetic fields, which directly influence the jet collimation and evolution.
Most recently, a prominent multi-band flare was reported in \cite{Carnerero15}, which peaked in 2013, spanning from radio wavelengths to $\gamma$-rays, followed by a prolonged period of quiescence \citep{McCall24}.
The delay between optical bands and $\gamma$-rays was marginal, whereas a two month delay was found between $\gamma$-rays and the millimetre radio band.
In parallel, the optical polarisation percentage rose sixfold during the multi-band flare onset \citep{Carnerero15}.
Here we explore the connection between this multi-band flare and structural changes in its parsec-scale jet, manifested in linear polarisation and electric vector position angle (EVPA) rotations, seeking a better understanding of the underlying driving mechanisms.
In Sect.~\ref{sec:Results} we briefly present the observations and expand on our results.
Section~\ref{sec:Discussion} we discuss a possible of the flaring event and in Sect.~\ref{sec:Conclusions} we summarise our findings.

\section{Methods}\label{sec:Results}
\subsection{Observations and data analysis}
For this work we utilised publicly available VLBI data obtained with the Very Long Baseline Array (VLBA) at 43\,GHz as part of the VLBA-BU-BLAZAR programme\footnote{\url{https://www.bu.edu/blazars/BEAM-ME.html}}.
A detailed description of the observations and how the data that we used were calibrated in total intensity and polarisation is reported in \cite{Jorstad17}.
During the time frame we review here, corresponding to modified Julian dates (MJD) between 56152 and 56349, \OJ\ was observed eight times.
For these epochs, we extracted total intensity, linear polarisation, and EVPA information using the geometrical model-fitting capability of the \texttt{eht-imaging} framework \citep{Chael16, Roelofs23}.
In a nutshell, geometrical model-fitting describes the structural and morphological properties of the observed emission, typically using simple parametric components such as Gaussian functions.
The advantage of this approach lies in the fact that the jet can be decomposed into such Gaussian components, for each of which all relevant measurables are identifiable. 
This makes it easier to track the connection between jet activity and light curve variability as manifested through multi-band flares.
Details about our modelling approach can be found in \cite{Paraschos24a} and \cite{Paraschos24c}.
The accompanying weekly binned $\gamma$-ray light curve was obtained from the Fermi Large Area Telescope Collaboration \citep[Fermi-LAT; see][]{Atwood09, Ajello20} repository\footnote{\url{https://fermi.gsfc.nasa.gov/ssc/data/access/lat/msl_lc/}
}, containing publicly available daily, weekly, and monthly binned data.

\subsection{Results}

\begin{figure}
\centering
\includegraphics[scale=0.5]{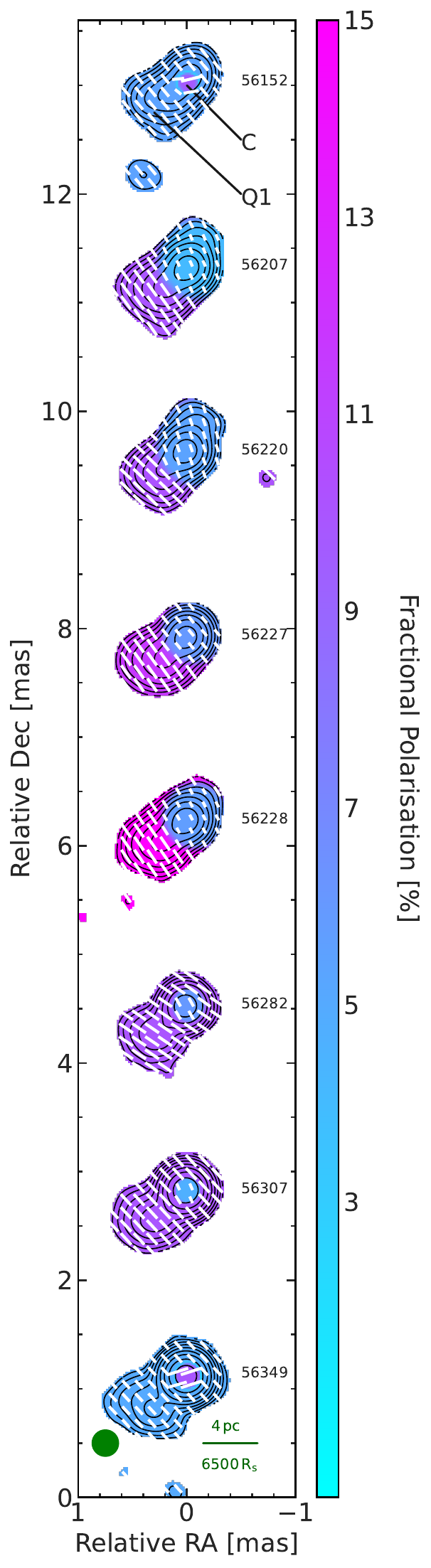}
  \caption{
    Stokes I image (contours) and fractional polarisation geometrical model-fit (colour) of \OJ\ showcasing the epochs close in time to the MJD 56152--56349 (year: 2012-2013) $\gamma$-ray flare.
    The colour scale is linear and the units are in \%.
    The dark green ellipse in the lower left corner shows the common convolving circular beam size of \(\beam\).
    The green bar (bottom right) corresponds to a projected distance of $6500 R_\textrm{S}$ (Schwarzschild radius).
    Superimposed are the EVPAs (white sticks), showcasing the direction of the electric field and by extension that of the magnetic field (perpendicular).
    Their length is proportional to the fractional polarisation values.
    The lowest Stokes I contour cutoﬀ is at $5\sigma_\textrm{I}$ was implemented, to only include high $S/N$ areas (with \(\sigma_\textrm{I} = \rms\)).
    The contour levels are at 0.25, 0.5, 1, 2, 4, 8, 16, 32, and 64\% of the total intensity peak per epoch, the MJDs of which are denoted next to each observation.
    The core C and component Q1 are denoted in the first epoch.
    The EVPAs in C start out along the bulk jet flow (southeast direction).
    As the shock front progresses downstream, they exhibit a clear turn, becoming perpendicular to the bulk jet flow and aligned to the EVPAs in Q1.
    }
   \label{fig:VLBI}
\end{figure}

Our analysis revealed that the \OJ\ jet is best approximated by two components in the time frame of interest, labelled as `C' (core, which we assumed to be stationary) and `Q1'.
In this framework C corresponds to the compact, optically thick region near the jet base and Q1 to the more extended jet emission.
Figure~\ref{fig:VLBI} displays the progression in the evolution of the overall jet morphology, with the parameters of the two components being listed in Table~\ref{table:Params}.
The parameters include the component identification, observation date, flux density (F0), full width at half maximum (FWHM) of each component, the position of Q1 relative to C, their fractional polarisation $m$, and their EVPAs.
Our results show that the extended jet (Q1) started out in a quiescent state (with $m_\textrm{C}>m_\textrm{Q1}$), then $m_\textrm{Q1}$ increased during the entirety of the $\gamma$-ray flare event duration (reaching a peak of $m_\textrm{Q1}\sim15\%$), before returning to its quiescent state again in the last epoch.
Simultaneously, $m_\textrm{C}\sim10\%$ before and after the flare but decreased to approximately half during it.
Interestingly, the EVPAs in C appear aligned to the bulk jet flow before and after the $\gamma$-ray flare event, while turning perpendicular to it during the event.
Furthermore, the overall flux density seems to increase during the event but then remains on higher levels.
Finally, the distance between Q1 and C is slowly increasing (\(\mu=\vel\), which is slower than the velocity of the source reported in the 1990s; see \citealt{Jorstad01}).

\begin{figure*}
\centering
\includegraphics[scale=0.75, trim={3.5cm, 0cm, 3.5cm, 0cm}]{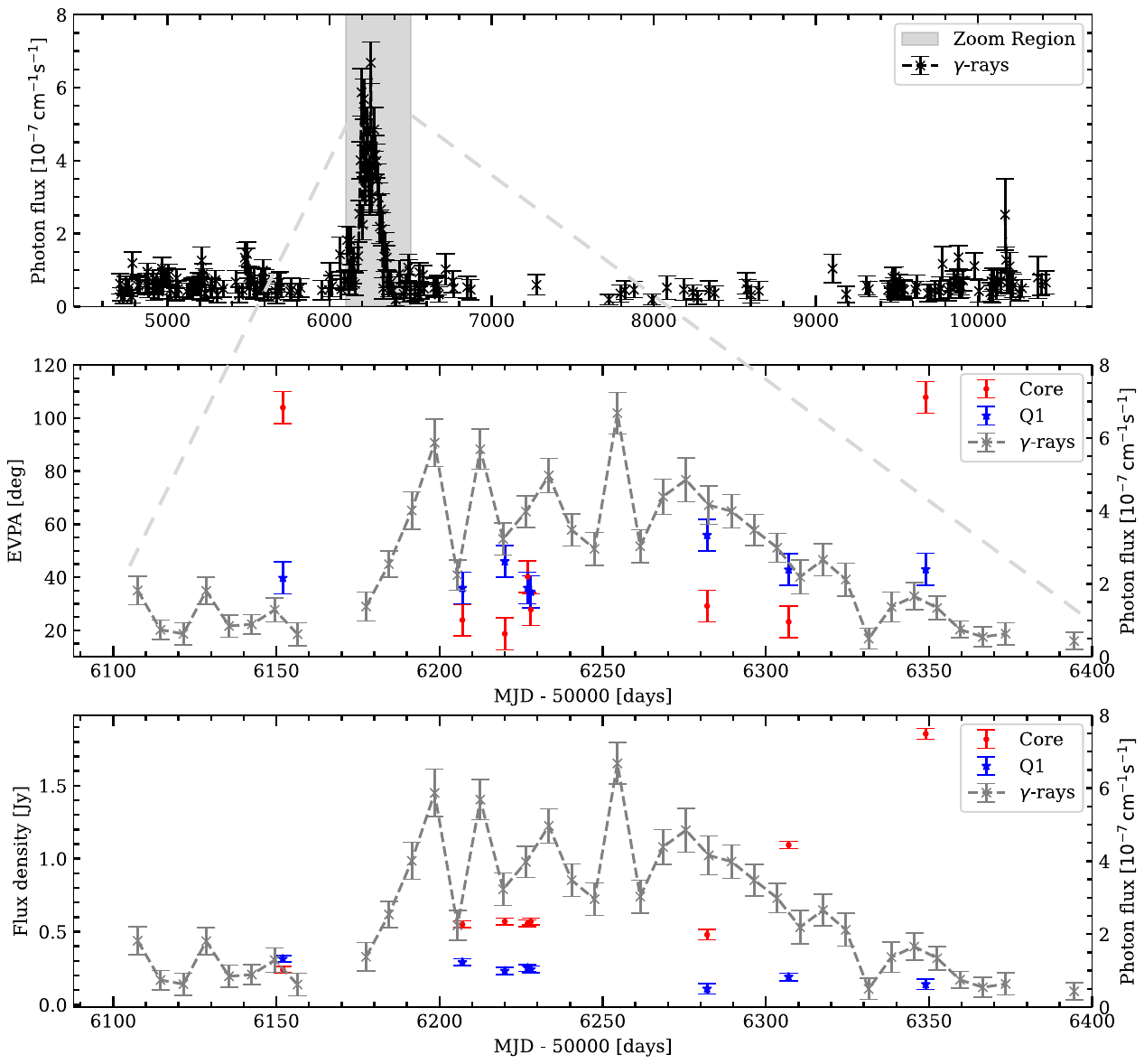}
  \caption{
    \OJ\ component EVPAs and flux density, and the Fermi $\gamma$-ray-LAT light curve.
    \emph{Top}: The grey-shaded area emphasises the period in question, when the prominent $\gamma$-ray flare occurred.
    In the years after, the source has remained quiescent.
    \emph{Middle}: The zoom-in of the $\gamma$-ray flare time frame (grey-shaded area of the top panel), revealing a multiple peaked profile. 
    Additionally, the EVPAs in C (red) and Q1 (blue) were perpendicular to each other immediately before and after the $\gamma$-ray flare, but then rotated to be aligned during the flaring event.
    \emph{Bottom}: Same zoom-in region as above with the flux density of C denoted in red and the one for Q1 in blue.
    An increasing trend in C is identified.
    }
     \label{fig:gamma}
\end{figure*}

Simultaneously, the $\gamma$-ray light-curve shown in the top panel of Fig.~\ref{fig:gamma} exhibited a very bright flare, the prominence of which has not been repeated since then.
In parallel, a multi-band flare was also reported in the optical, near IR, and millimetre radio bands by \cite{Carnerero15}.
The authors also found an increase in the polarised flux and rotations in the optical EVPAs (although a clear trend was lacking).
In the middle and bottom panels of Fig.~\ref{fig:gamma} we over-plotted the radio wavelength EVPAs and flux density of C and Q1 during the flaring event.
We can clearly observe that the $\gamma$-ray flare is accompanied by an EVPA rotation in the radio as well, accompanied by a flux increase.

\begin{table*}
\caption{Summary of the component parameters}
\label{table:Params}
\centering
\begin{tabular}{ccccccc}
ID & Obs. [MJD - 50000] & F0 [Jy] & FWHM [mas] & Position [mas, mas] & Frac. pol [\%] & EVPA [deg]\\
\hline\hline
 C  &  6152 &  $0.240\pm0.036$ &  $0.03\pm0.048$ &  [0.00, 0.00]  &  $9.9\pm1.5$  &  $104\pm6$\\
 C  &  6207 &  $0.550\pm0.083$ &  $0.10\pm0.048$ &  [0.00, 0.00]  &  $4.1\pm0.6$  &  $24\pm6$ \\
 C  &  6220 &  $0.570\pm0.085$ &  $0.09\pm0.048$ &  [0.00, 0.00]  &  $5.3\pm0.8$  &  $19\pm6$ \\
 C  &  6227 &  $0.558\pm0.084$ &  $0.08\pm0.048$ &  [0.00, 0.00]  &  $6.0\pm0.9$  &  $40\pm6$ \\
 C  &  6228 &  $0.570\pm0.085$ &  $0.08\pm0.048$ &  [0.00, 0.00]  &  $5.7\pm0.9$  &  $28\pm6$ \\
 C  &  6282 &  $0.479\pm0.072$ &  $0.04\pm0.048$ &  [0.00, 0.00]  &  $6.0\pm0.9$  &  $29\pm6$ \\
 C  &  6307 &  $1.094\pm0.164$ &  $0.02\pm0.048$ &  [0.00, 0.00]  &  $4.7\pm0.9$  &  $23\pm6$ \\
 C  &  6349 &  $1.855\pm0.278$ &  $0.03\pm0.048$ &  [0.00, 0.00]  &  $10.0\pm3.9$ &  $108\pm6$\\
 Q1 &  6152 &  $0.314\pm0.047$ &  $0.20\pm0.048$ &  [0.15, -0.08] &  $5.4\pm0.8$  &  $40\pm6$ \\
 Q1 &  6207 &  $0.290\pm0.043$ &  $0.21\pm0.048$ &  [0.21, -0.16] &  $10.0\pm1.5$ &  $36\pm6$ \\
 Q1 &  6220 &  $0.231\pm0.035$ &  $0.19\pm0.048$ &  [0.22, -0.16] &  $10.1\pm1.5$ &  $46\pm6$ \\ 
 Q1 &  6227 &  $0.251\pm0.038$ &  $0.19\pm0.048$ &  [0.21, -0.17] &  $11.6\pm1.7$ &  $36\pm6$ \\
 Q1 &  6228 &  $0.242\pm0.036$ &  $0.19\pm0.048$ &  [0.22, -0.15] &  $15.0\pm2.3$ &  $34\pm6$ \\
 Q1 &  6282 &  $0.109\pm0.016$ &  $0.19\pm0.048$ &  [0.20, -0.14] &  $10.0\pm1.5$ &  $56\pm6$ \\
 Q1 &  6307 &  $0.190\pm0.029$ &  $0.19\pm0.048$ &  [0.22, -0.22] &  $10.0\pm1.5$ &  $43\pm6$ \\
 Q1 &  6349 &  $0.140\pm0.021$ &  $0.22\pm0.048$ &  [0.23, -0.02] &  $5.0\pm0.8$  &  $43\pm6$ \\
\hline
\end{tabular}
\end{table*}

\section{Discussion}\label{sec:Discussion}

The true nature of $\gamma$-ray flares is one of the fundamental open questions related to jet activity.
In order to produce such a flare, a region of internal photons must be present for synchrotron self-Compton \citep{Konigl81} scattering (inverse Compton-scattering) to occur; the so-called blazar zone \citep[see, e.g.][for a review]{Hovatta19b}.
The exact location of this flare origin region with regard to the radio core observed with VLBI, remains unclear. 
Both upstream (near the central engine) and downstream (extended jet) locations of the VLBI core have recently been reported as the possible origin of flares and this includes blazars \citep[e.g.][]{Rani18} and, interestingly, radio galaxies as well \citep[e.g.][]{Paraschos23}, with the latter also showing evidence of a connection to jet feature ejections \citep{Agudo13, Paraschos22}.
Furthermore, jets could be transversely stratified, for example exhibiting a so-called `spine-sheath' geometry, where a fast inner plasma flow is surrounded by a slower moving one \citep{Sol89, Laing96, Paraschos24b}.
In this case, the blazar zone might be connected to either the spine or the sheath \citep[see, e.g.][for a review]{Blandford19a}.

A common strategy of localising the blazar zone is to study the correlation between the onset and duration of a multi-band flare at the different frequency bands.
Interestingly, as discussed in \cite{Carnerero15}, the \OJ\ multi-band light curves recorded in $\gamma$-rays, optical, IR, and radio bands have exhibited a significant correlation over time, when taking into account the source's past main flaring events.
Such a manifestation is commonplace for blazars like \OJ\ \citep{LeonTavares11}.
The simultaneity of this multi-band flare and the activity in the jet dynamics found in our work strongly indicates a connection between them.
A natural explanation of this phenomenon is that the blazar region is optically thin to radio wavelengths.
The increase of the fractional polarisation in Q1 and the EVPAs' rotation in C can be explained by a moving shock front, as reported, for example, in \cite{Liodakis22}.
This shock front then appears to be ripping through a standing recollimation shock in the vicinity of the VLBI core \citep{Daly88}, resulting also in the observed peak multiplicity of the $\gamma$-rays (see middle and bottom panel of Fig.~\ref{fig:gamma}).
Such activity has also been reported for the case of PKS\,1510–089 in \cite{Hess21}.
We note with interest, that a similar such morphology has also been reported in other sources \citep[e.g.][]{Larionov13, Liodakis20}.

At this point we can additionally consider two alternative scenarios for the observed behaviour, namely either a kink instability \citep{Nalewajko17}, or geometric effects.
For the former we would expect a decrease in optical polarisation, which is not reported in \cite{Carnerero15}.
A swinging nozzle model, in which abrupt changes to the viewing angle result in the flux density increase in all bands, has been proposed by \cite{Jorstad04b} to explain the latter scenario for \OJ.
The expected accompanying intranight variability, however, has been lacking \citep[as reported in][]{Marchesini16}.

Another ansatz is the commonly invoked shock-in-jet model \cite{Marscher85}, which offers a convincing explanation of the increase in optical polarisation, as well as in fractional polarisation in the time frame of the $\gamma$-ray flare.
In that case, the travelling shock front is expected to compress the magnetic field component perpendicular to the outflow and to cause a rotation of the optical EVPAs \citep{Blinov18}.
However, similarly to the case of the radio sources 3C\,454.3 \citep{Liodakis20} and S4\,0954+65 \citep{Kouch24}, in \OJ\ the EVPAs are the ones directed perpendicular to jet flow direction, which means that the magnetic field lines are parallel to it.
Therefore, with the addition of the case of \OJ\ to the above two, there is now mounting evidence of the scenario described in \cite{Marscher02}.
In that work, the authors postulate that ambient magnetic field lines parallel to the jet axis are possibly amplified by shearing processes or interactions with the surrounding medium.
A passing shock front would then result in increased particle acceleration, which leads to the observed geometry of the EVPAs in C.
We propose here that this description fits the observed phenomenology; the shock propagates through and beyond C, which harbours an area of magnetic field lines parallel to the bulk jet flow.
Particles are accelerated and then cool down, causing the $\gamma$-ray flare in this region of magnetic field lines streamlined to the jet axis.
Furthermore, in this interpretation the magnetic field is not amplified (no compression to amplify the magnetic flux takes place), explaining the lower fractional polarisation values of C during the $\gamma$-ray flare.
Finally, at the decay phase of the $\gamma$-ray flare, the magnetic field that the shock interacts with becomes more turbulent.
This results in the compression of the field, providing order and restoring the EVPA orientation parallel to the bulk jet flow \citep[see][]{Kouch24}, which explains the observed rise of fractional polarisation in C after the flare.

Overall, the rotation of the EVPAs in the central region C being simultaneous with the increase in polarisation of the jet (Q1), is evidence in favour of the blazar region of \OJ\ being in the vicinity of the 43\,GHz VLBI core.
In our present understanding of blazar jets \citep[e.g.][]{Marscher08} this corresponds to a $\sim$parsec-scale distance to central engine.
Below, we also consider sub-parsec scale distances for the blazar zone.
Specifically, the dusty torus, a possible source of seed electrons needed to cause the $\gamma$-ray flare \citep{Saito15, Ahnen17}, would be located too far upstream, close to the central engine.
Similarly, a blazar region supported by photons from the broad line region, as suggested by \cite{Ghisellini96}, would require the $\gamma$-ray flare to lag the optical one, which is not conclusively observed in \cite{Carnerero15}.
Finally, if the blazar zone is located even further downstream (few parsec away from the central engine), as indicated by the evolution of the jet seen in Fig.~\ref{fig:VLBI}, the origin of the seed photons could be within a surrounding jet sheath \citep[e.g.][]{Attridge99, MacDonald15}. 
This scenario, which particularly works for FSRQs such as \OJ, has been reported to be able to produce the observed $\gamma$-ray flares \citep{MacDonald17}, albeit in a suppressed magnetic field strength regime.
In that case, we can also make a prediction about future $\gamma$-ray flares in the source, based on the work presented by \cite{Blinov21}.
Specifically, the authors there show that condensation rings within the jet sheath, surrounding the inner spine, can provide the necessary seed photons.
Under these model assumptions, future $\gamma$-ray flares are predicted to exhibit a similar pattern to the one found here (i.e. multiple peaks).

\section{Conclusions} \label{sec:Conclusions}

In this work we investigated the connection between the polarisation properties of the \OJ\ jet and a prominent $\gamma$-ray flare detected in the same source.
Our findings can be summarised as follows:
\begin{itemize}
    \item The \OJ\ jet exhibits remarkable morphological changes through the course of the flaring event, detected in $\gamma$-rays, optical, near infrared, and radio bands, as well as in polarised emission.
    Specifically, the VLBI fractional linear polarisation of its extended emission (component Q1) is heightened during the $\gamma$-ray outburst, while being in lower states immediately before and after.
    The EVPAs appear to also be affected; during the $\gamma$-ray flare they are perpendicular to the bulk jet flow, whereas before and after they are not.
    \item This behaviour is consistent with a moving shock front, travelling through a recollimation shock, while causing the photons to be inverse Compton upscattered into creating the $\gamma$-ray flare.
\end{itemize}

Our work presented here showcases the importance of monitoring blazars both with VLBI and with individual telescopes at multiple bands, in order to gain insights into their underlying physical mechanisms.
Future, higher sensitivity instruments, such as the next-generation Very Large Array (ngVLA) will enable us to image fainter sources, such as \OJ, in greater detail and thus leverage the possibilities that multi-messenger astronomy has ushered in.

\begin{acknowledgements}
      We thank I.~Liodakis and an anonymous referee for their valuable comments which greatly improved this manuscript and L.~Debbrecht for the insightful discussions.
      This research is supported by the European Research Council advanced grant “M2FINDERS - Mapping Magnetic Fields with INterferometry Down to Event hoRizon Scales” (Grant No. 101018682). 
      This study makes use of VLBA data from the VLBA-BU Blazar Monitoring Program (BEAM-ME and VLBA-BU-BLAZAR;
      \url{http://www.bu.edu/blazars/BEAM-ME.html}), funded by NASA through the Fermi Guest Investigator Program. The VLBA is an instrument of the National Radio Astronomy Observatory. The National Radio Astronomy Observatory is a facility of the National Science Foundation operated by Associated Universities, Inc.
      This research has made use of the NASA/IPAC Extragalactic Database (NED), which is operated by the Jet Propulsion Laboratory, California Institute of Technology, under contract with the National Aeronautics and Space Administration. 
      This research has also made use of NASA's Astrophysics Data System Bibliographic Services. 
      Finally, this research made use of the following python packages: {\it numpy} \citep{Harris20}, {\it scipy} \citep{2020SciPy-NMeth}, {\it matplotlib} \citep{Hunter07}, {\it astropy} \citep{2013A&A...558A..33A, 2018AJ....156..123A} and {\it Uncertainties: a Python package for calculations with uncertainties.
      }
\end{acknowledgements}

%
%

\bibliographystyle{aa} 
\bibliography{aanda} 

\begin{thebibliography}{56}
\expandafter\ifx\csname natexlab\endcsname\relax\def\natexlab#1{#1}\fi

\bibitem[{{Abdo} {et~al.}(2010){Abdo}, {Ackermann}, {Ajello}, {Antolini},
  {Baldini}, {Ballet}, {Barbiellini}, {Baring}, {Bastieri}, {Bechtol},
  {Bellazzini}, {Berenji}, {Blandford}, {Bloom}, {Bonamente}, {Borgland},
  {Bregeon}, {Brez}, {Brigida}, {Bruel}, {Buehler}, {Buson}, {Caliandro},
  {Cameron}, {Carrigan}, {Casandjian}, {Cavazzuti}, {Cecchi}, {{\c{C}}elik},
  {Chekhtman}, {Chen}, {Chiang}, {Ciprini}, {Claus}, {Cohen-Tanugi},
  {Colafrancesco}, {Conrad}, {Cutini}, {Dermer}, {de Palma}, {Digel}, {Silva},
  {Drell}, {Dubois}, {Dumora}, {Farnier}, {Favuzzi}, {Fegan}, {Ferrara},
  {Focke}, {Frailis}, {Fukazawa}, {Fusco}, {Gargano}, {Gasparrini}, {Gehrels},
  {Giebels}, {Giglietto}, {Giommi}, {Giordano}, {Giroletti}, {Glanzman},
  {Godfrey}, {Grandi}, {Grenier}, {Guillemot}, {Guiriec}, {Hadasch}, {Harding},
  {Hayashida}, {Horan}, {Hughes}, {Itoh}, {Jackson}, {J{\'o}hannesson},
  {Johnson}, {Johnson}, {Kamae}, {Katagiri}, {Kataoka}, {Kawai},
  {Kn{\"o}dlseder}, {Kuss}, {Lande}, {Latronico}, {Longo}, {Loparco}, {Lott},
  {Lovellette}, {Lubrano}, {Madejski}, {Makeev}, {Mazziotta}, {McEnery},
  {McGlynn}, {Meurer}, {Michelson}, {Mitthumsiri}, {Mizuno}, {Monte},
  {Monzani}, {Morselli}, {Moskalenko}, {Murgia}, {Nestoras}, {Nolan}, {Norris},
  {Nuss}, {Ohsugi}, {Okumura}, {Orlando}, {Ormes}, {Ozaki}, {Paneque},
  {Panetta}, {Parent}, {Pelassa}, {Pepe}, {Pesce-Rollins}, {Piron}, {Porter},
  {Rain{\`o}}, {Rando}, {Razzano}, {Reimer}, {Reimer}, {Reyes}, {Rodriguez},
  {Roth}, {Ryde}, {Sadrozinski}, {Sambruna}, {Sander}, {Sato}, {Sgr{\`o}},
  {Shaw}, {Siskind}, {Smith}, {Spandre}, {Spinelli}, {Stawarz}, {Stecker},
  {Strickman}, {Suson}, {Takahashi}, {Takahashi}, {Tanaka}, {Thayer}, {Thayer},
  {Thompson}, {Tibolla}, {Torres}, {Tosti}, {Tramacere}, {Uchiyama}, {Usher},
  {Vasileiou}, {Vilchez}, {Villata}, {Vitale}, {von Kienlin}, {Waite}, {Wang},
  {Winer}, {Wood}, {Yang}, {Ylinen}, {Ziegler}, {Tavecchio}, {Sikora},
  {Schady}, {Roming}, {Chester}, \& {Maraschi}}]{Abdo10}
{Abdo}, A.~A., {Ackermann}, M., {Ajello}, M., {et~al.} 2010, \apj, 716, 835

\bibitem[{{Agudo}(2013)}]{Agudo13}
{Agudo}, I. 2013, in European Physical Journal Web of Conferences, Vol.~61,
  European Physical Journal Web of Conferences, 04002

\bibitem[{{Ahnen} {et~al.}(2017){Ahnen}, {Ansoldi}, {Antonelli}, {Arcaro},
  {Babi{\'c}}, {Banerjee}, {Bangale}, {Barres de Almeida}, {Barrio}, {Becerra
  Gonz{\'a}lez}, {Bednarek}, {Bernardini}, {Berti}, {Biasuzzi}, {Biland},
  {Blanch}, {Bonnefoy}, {Bonnoli}, {Borracci}, {Bretz}, {Carosi}, {Carosi},
  {Chatterjee}, {Colin}, {Colombo}, {Contreras}, {Cortina}, {Covino}, {Cumani},
  {Da Vela}, {Dazzi}, {De Angelis}, {De Lotto}, {de O{\~n}a Wilhelmi}, {Di
  Pierro}, {Doert}, {Dom{\'\i}nguez}, {Dominis Prester}, {Dorner}, {Doro},
  {Einecke}, {Eisenacher Glawion}, {Elsaesser}, {Engelkemeier}, {Fallah
  Ramazani}, {Fern{\'a}ndez-Barral}, {Fidalgo}, {Fonseca}, {Font}, {Fruck},
  {Galindo}, {Garc{\'\i}a L{\'o}pez}, {Garczarczyk}, {Gaug}, {Giammaria},
  {Godinovi{\'c}}, {Gora}, {Guberman}, {Hadasch}, {Hahn}, {Hassan},
  {Hayashida}, {Herrera}, {Hose}, {Hrupec}, {Hughes}, {Idec}, {Ishio},
  {Kodani}, {Konno}, {Kubo}, {Kushida}, {Lelas}, {Lindfors}, {Lombardi},
  {Longo}, {L{\'o}pez}, {Majumdar}, {Makariev}, {Mallot}, {Maneva},
  {Manganaro}, {Mannheim}, {Maraschi}, {Mariotti}, {Mart{\'\i}nez}, {Mazin},
  {Menzel}, {Mirzoyan}, {Moralejo}, {Moretti}, {Nakajima}, {Neustroev},
  {Niedzwiecki}, {Nievas Rosillo}, {Nilsson}, {Nishijima}, {Noda},
  {Nogu{\'e}s}, {N{\"o}the}, {Paiano}, {Palacio}, {Palatiello}, {Paneque},
  {Paoletti}, {Paredes}, {Paredes-Fortuny}, {Pedaletti}, {Peresano}, {Perri},
  {Persic}, {Poutanen}, {Prada Moroni}, {Prandini}, {Puljak}, {Garcia},
  {Reichardt}, {Rhode}, {Rib{\'o}}, {Rico}, {Saito}, {Satalecka}, {Schroeder},
  {Schweizer}, {Shore}, {Sillanp{\"a}{\"a}}, {Sitarek}, {Snidaric},
  {Sobczynska}, {Stamerra}, {Strzys}, {Suri{\'c}}, {Takalo}, {Tavecchio},
  {Temnikov}, {Terzi{\'c}}, {Tescaro}, {Teshima}, {Torres}, {Torres-Alb{\`a}},
  {Toyama}, {Treves}, {Vanzo}, {Vazquez Acosta}, {Vovk}, {Ward}, {Will}, {Wu},
  {Krau{\ss}}, {Schulz}, {Kadler}, {Wilms}, {Ros}, {Bach}, {Beuchert},
  {Langejahn}, {Wendel}, {Gehrels}, {Baumgartner}, {Markwardt}, {M{\"u}ller},
  {Grinberg}, {Hovatta}, \& {Magill}}]{Ahnen17}
{Ahnen}, M.~L., {Ansoldi}, S., {Antonelli}, L.~A., {et~al.} 2017, \aap, 603,
  A25

\bibitem[{{Ajello} {et~al.}(2020){Ajello}, {Angioni}, {Axelsson}, {Ballet},
  {Barbiellini}, {Bastieri}, {Becerra Gonzalez}, {Bellazzini}, {Bissaldi},
  {Bloom}, {Bonino}, {Bottacini}, {Bruel}, {Buson}, {Cafardo}, {Cameron},
  {Cavazzuti}, {Chen}, {Cheung}, {Ciprini}, {Costantin}, {Cutini}, {D'Ammando},
  {de la Torre Luque}, {de Menezes}, {de Palma}, {Desai}, {Di Lalla}, {Di
  Venere}, {Dom{\'\i}nguez}, {Dirirsa}, {Ferrara}, {Finke}, {Franckowiak},
  {Fukazawa}, {Funk}, {Fusco}, {Gargano}, {Garrappa}, {Gasparrini},
  {Giglietto}, {Giordano}, {Giroletti}, {Green}, {Grenier}, {Guiriec},
  {Harita}, {Hays}, {Horan}, {Itoh}, {J{\'o}hannesson}, {Kovac'evic'},
  {Krauss}, {Kreter}, {Kuss}, {Larsson}, {Leto}, {Li}, {Liodakis}, {Longo},
  {Loparco}, {Lott}, {Lovellette}, {Lubrano}, {Madejski}, {Maldera},
  {Manfreda}, {Mart{\'\i}-Devesa}, {Massaro}, {Mazziotta}, {Mereu}, {Meyer},
  {Migliori}, {Mirabal}, {Mizuno}, {Monzani}, {Morselli}, {Moskalenko},
  {Negro}, {Nemmen}, {Nuss}, {Ojha}, {Ojha}, {Omodei}, {Orienti}, {Orlando},
  {Ormes}, {Paliya}, {Pei}, {Pe{\~n}a-Herazo}, {Persic}, {Pesce-Rollins},
  {Petrov}, {Piron}, {Poon}, {Principe}, {Rain{\`o}}, {Rando}, {Rani},
  {Razzano}, {Razzaque}, {Reimer}, {Reimer}, {Schinzel}, {Serini}, {Sgr{\`o}},
  {Siskind}, {Spandre}, {Spinelli}, {Suson}, {Tachibana}, {Thompson}, {Torres},
  {Torresi}, {Troja}, {Valverde}, {van Zyl}, \& {Yassine}}]{Ajello20}
{Ajello}, M., {Angioni}, R., {Axelsson}, M., {et~al.} 2020, \apj, 892, 105

\bibitem[{{Astropy Collaboration} {et~al.}(2018){Astropy Collaboration},
  {Price-Whelan}, {Sip{\H{o}}cz}, {G{\"u}nther}, {Lim}, {Crawford}, {Conseil},
  {Shupe}, {Craig}, {Dencheva}, {Ginsburg}, {VanderPlas}, {Bradley},
  {P{\'e}rez-Su{\'a}rez}, {de Val-Borro}, {Aldcroft}, {Cruz}, {Robitaille},
  {Tollerud}, {Ardelean}, {Babej}, {Bach}, {Bachetti}, {Bakanov}, {Bamford},
  {Barentsen}, {Barmby}, {Baumbach}, {Berry}, {Biscani}, {Boquien}, {Bostroem},
  {Bouma}, {Brammer}, {Bray}, {Breytenbach}, {Buddelmeijer}, {Burke},
  {Calderone}, {Cano Rodr{\'\i}guez}, {Cara}, {Cardoso}, {Cheedella}, {Copin},
  {Corrales}, {Crichton}, {D'Avella}, {Deil}, {Depagne}, {Dietrich}, {Donath},
  {Droettboom}, {Earl}, {Erben}, {Fabbro}, {Ferreira}, {Finethy}, {Fox},
  {Garrison}, {Gibbons}, {Goldstein}, {Gommers}, {Greco}, {Greenfield},
  {Groener}, {Grollier}, {Hagen}, {Hirst}, {Homeier}, {Horton}, {Hosseinzadeh},
  {Hu}, {Hunkeler}, {Ivezi{\'c}}, {Jain}, {Jenness}, {Kanarek}, {Kendrew},
  {Kern}, {Kerzendorf}, {Khvalko}, {King}, {Kirkby}, {Kulkarni}, {Kumar},
  {Lee}, {Lenz}, {Littlefair}, {Ma}, {Macleod}, {Mastropietro}, {McCully},
  {Montagnac}, {Morris}, {Mueller}, {Mumford}, {Muna}, {Murphy}, {Nelson},
  {Nguyen}, {Ninan}, {N{\"o}the}, {Ogaz}, {Oh}, {Parejko}, {Parley}, {Pascual},
  {Patil}, {Patil}, {Plunkett}, {Prochaska}, {Rastogi}, {Reddy Janga},
  {Sabater}, {Sakurikar}, {Seifert}, {Sherbert}, {Sherwood-Taylor}, {Shih},
  {Sick}, {Silbiger}, {Singanamalla}, {Singer}, {Sladen}, {Sooley},
  {Sornarajah}, {Streicher}, {Teuben}, {Thomas}, {Tremblay}, {Turner},
  {Terr{\'o}n}, {van Kerkwijk}, {de la Vega}, {Watkins}, {Weaver}, {Whitmore},
  {Woillez}, {Zabalza}, \& {Astropy Contributors}}]{2018AJ....156..123A}
{Astropy Collaboration}, {Price-Whelan}, A.~M., {Sip{\H{o}}cz}, B.~M., {et~al.}
  2018, \aj, 156, 123

\bibitem[{{Astropy Collaboration} {et~al.}(2013){Astropy Collaboration},
  {Robitaille}, {Tollerud}, {Greenfield}, {Droettboom}, {Bray}, {Aldcroft},
  {Davis}, {Ginsburg}, {Price-Whelan}, {Kerzendorf}, {Conley}, {Crighton},
  {Barbary}, {Muna}, {Ferguson}, {Grollier}, {Parikh}, {Nair}, {Unther},
  {Deil}, {Woillez}, {Conseil}, {Kramer}, {Turner}, {Singer}, {Fox}, {Weaver},
  {Zabalza}, {Edwards}, {Azalee Bostroem}, {Burke}, {Casey}, {Crawford},
  {Dencheva}, {Ely}, {Jenness}, {Labrie}, {Lim}, {Pierfederici}, {Pontzen},
  {Ptak}, {Refsdal}, {Servillat}, \& {Streicher}}]{2013A&A...558A..33A}
{Astropy Collaboration}, {Robitaille}, T.~P., {Tollerud}, E.~J., {et~al.} 2013,
  \aap, 558, A33

\bibitem[{{Attridge} {et~al.}(1999){Attridge}, {Roberts}, \&
  {Wardle}}]{Attridge99}
{Attridge}, J.~M., {Roberts}, D.~H., \& {Wardle}, J. F.~C. 1999, \apjl, 518,
  L87

\bibitem[{{Atwood} {et~al.}(2009){Atwood}, {Abdo}, {Ackermann}, {Althouse},
  {Anderson}, {Axelsson}, {Baldini}, {Ballet}, {Band}, {Barbiellini},
  {Bartelt}, {Bastieri}, {Baughman}, {Bechtol}, {B{\'e}d{\'e}r{\`e}de},
  {Bellardi}, {Bellazzini}, {Berenji}, {Bignami}, {Bisello}, {Bissaldi},
  {Blandford}, {Bloom}, {Bogart}, {Bonamente}, {Bonnell}, {Borgland},
  {Bouvier}, {Bregeon}, {Brez}, {Brigida}, {Bruel}, {Burnett}, {Busetto},
  {Caliandro}, {Cameron}, {Caraveo}, {Carius}, {Carlson}, {Casandjian},
  {Cavazzuti}, {Ceccanti}, {Cecchi}, {Charles}, {Chekhtman}, {Cheung},
  {Chiang}, {Chipaux}, {Cillis}, {Ciprini}, {Claus}, {Cohen-Tanugi},
  {Condamoor}, {Conrad}, {Corbet}, {Corucci}, {Costamante}, {Cutini}, {Davis},
  {Decotigny}, {DeKlotz}, {Dermer}, {de Angelis}, {Digel}, {do Couto e Silva},
  {Drell}, {Dubois}, {Dumora}, {Edmonds}, {Fabiani}, {Farnier}, {Favuzzi},
  {Flath}, {Fleury}, {Focke}, {Funk}, {Fusco}, {Gargano}, {Gasparrini},
  {Gehrels}, {Gentit}, {Germani}, {Giebels}, {Giglietto}, {Giommi}, {Giordano},
  {Glanzman}, {Godfrey}, {Grenier}, {Grondin}, {Grove}, {Guillemot}, {Guiriec},
  {Haller}, {Harding}, {Hart}, {Hays}, {Healey}, {Hirayama}, {Hjalmarsdotter},
  {Horn}, {Hughes}, {J{\'o}hannesson}, {Johansson}, {Johnson}, {Johnson},
  {Johnson}, {Johnson}, {Kamae}, {Katagiri}, {Kataoka}, {Kavelaars}, {Kawai},
  {Kelly}, {Kerr}, {Klamra}, {Kn{\"o}dlseder}, {Kocian}, {Komin}, {Kuehn},
  {Kuss}, {Landriu}, {Latronico}, {Lee}, {Lee}, {Lemoine-Goumard}, {Lionetto},
  {Longo}, {Loparco}, {Lott}, {Lovellette}, {Lubrano}, {Madejski}, {Makeev},
  {Marangelli}, {Massai}, {Mazziotta}, {McEnery}, {Menon}, {Meurer},
  {Michelson}, {Minuti}, {Mirizzi}, {Mitthumsiri}, {Mizuno}, {Moiseev},
  {Monte}, {Monzani}, {Moretti}, {Morselli}, {Moskalenko}, {Murgia},
  {Nakamori}, {Nishino}, {Nolan}, {Norris}, {Nuss}, {Ohno}, {Ohsugi}, {Omodei},
  {Orlando}, {Ormes}, {Paccagnella}, {Paneque}, {Panetta}, {Parent}, {Pearce},
  {Pepe}, {Perazzo}, {Pesce-Rollins}, {Picozza}, {Pieri}, {Pinchera}, {Piron},
  {Porter}, {Poupard}, {Rain{\`o}}, {Rando}, {Rapposelli}, {Razzano}, {Reimer},
  {Reimer}, {Reposeur}, {Reyes}, {Ritz}, {Rochester}, {Rodriguez}, {Romani},
  {Roth}, {Russell}, {Ryde}, {Sabatini}, {Sadrozinski}, {Sanchez}, {Sander},
  {Sapozhnikov}, {Parkinson}, {Scargle}, {Schalk}, {Scolieri}, {Sgr{\`o}},
  {Share}, {Shaw}, {Shimokawabe}, {Shrader}, {Sierpowska-Bartosik}, {Siskind},
  {Smith}, {Smith}, {Spandre}, {Spinelli}, {Starck}, {Stephens}, {Strickman},
  {Strong}, {Suson}, {Tajima}, {Takahashi}, {Takahashi}, {Tanaka}, {Tenze},
  {Tether}, {Thayer}, {Thayer}, {Thompson}, {Tibaldo}, {Tibolla}, {Torres},
  {Tosti}, {Tramacere}, {Turri}, {Usher}, {Vilchez}, {Vitale}, {Wang},
  {Watters}, {Winer}, {Wood}, {Ylinen}, \& {Ziegler}}]{Atwood09}
{Atwood}, W.~B., {Abdo}, A.~A., {Ackermann}, M., {et~al.} 2009, \apj, 697, 1071

\bibitem[{{Blandford} {et~al.}(2019){Blandford}, {Meier}, \&
  {Readhead}}]{Blandford19a}
{Blandford}, R., {Meier}, D., \& {Readhead}, A. 2019, \araa, 57, 467

\bibitem[{{Blinov} {et~al.}(2021){Blinov}, {Jorstad}, {Larionov}, {MacDonald},
  {Grishina}, {Kopatskaya}, {Larionova}, {Larionova}, {Morozova}, {Nikiforova},
  {Savchenko}, {Troitskaya}, \& {Troitsky}}]{Blinov21}
{Blinov}, D., {Jorstad}, S.~G., {Larionov}, V.~M., {et~al.} 2021, \mnras, 505,
  4616

\bibitem[{{Blinov} {et~al.}(2018){Blinov}, {Pavlidou}, {Papadakis},
  {Kiehlmann}, {Liodakis}, {Panopoulou}, {Angelakis}, {Balokovi{\'c}},
  {Hovatta}, {King}, {Kus}, {Kylafis}, {Mahabal}, {Maharana}, {Myserlis},
  {Paleologou}, {Papamastorakis}, {Pazderski}, {Pearson}, {Ramaprakash},
  {Readhead}, {Reig}, {Tassis}, \& {Zensus}}]{Blinov18}
{Blinov}, D., {Pavlidou}, V., {Papadakis}, I., {et~al.} 2018, \mnras, 474, 1296

\bibitem[{{Carnerero} {et~al.}(2015){Carnerero}, {Raiteri}, {Villata},
  {Acosta-Pulido}, {D'Ammando}, {Smith}, {Larionov}, {Agudo}, {Ar{\'e}valo},
  {Arkharov}, {Bach}, {Bachev}, {Ben{\'\i}tez}, {Blinov}, {Bozhilov}, {Buemi},
  {Bueno Bueno}, {Carosati}, {Casadio}, {Chen}, {Damljanovic}, {di Paola},
  {Efimova}, {Ehgamberdiev}, {Giroletti}, {G{\'o}mez}, {Gonz{\'a}lez-Morales},
  {Grinon-Marin}, {Grishina}, {Gurwell}, {Hiriart}, {Hsiao}, {Ibryamov},
  {Jorstad}, {Joshi}, {Kopatskaya}, {Kurtanidze}, {Kurtanidze},
  {L{\"a}hteenm{\"a}ki}, {Larionova}, {Larionova}, {L{\'a}zaro}, {Leto}, {Lin},
  {Lin}, {Manilla-Robles}, {Marscher}, {McHardy}, {Metodieva}, {Mirzaqulov},
  {Mokrushina}, {Molina}, {Morozova}, {Nikolashvili}, {Orienti}, {Ovcharov},
  {Panwar}, {Pastor Yabar}, {Puerto Gim{\'e}nez}, {Ramakrishnan}, {Richter},
  {Rossini}, {Sigua}, {Strigachev}, {Taylor}, {Tornikoski}, {Trigilio},
  {Troitskaya}, {Troitsky}, {Umana}, {Valcheva}, {Velasco}, {Vince}, {Wehrle},
  \& {Wiesemeyer}}]{Carnerero15}
{Carnerero}, M.~I., {Raiteri}, C.~M., {Villata}, M., {et~al.} 2015, \mnras,
  450, 2677

\bibitem[{{Chael} {et~al.}(2016){Chael}, {Johnson}, {Narayan}, {Doeleman},
  {Wardle}, \& {Bouman}}]{Chael16}
{Chael}, A.~A., {Johnson}, M.~D., {Narayan}, R., {et~al.} 2016, \apj, 829, 11

\bibitem[{{Daly} \& {Marscher}(1988)}]{Daly88}
{Daly}, R.~A. \& {Marscher}, A.~P. 1988, \apj, 334, 539

\bibitem[{{Enya} {et~al.}(2002){Enya}, {Yoshii}, {Kobayashi}, {Minezaki},
  {Suganuma}, {Tomita}, \& {Peterson}}]{Enya02}
{Enya}, K., {Yoshii}, Y., {Kobayashi}, Y., {et~al.} 2002, \apjs, 141, 31

\bibitem[{{Ghisellini} \& {Madau}(1996)}]{Ghisellini96}
{Ghisellini}, G. \& {Madau}, P. 1996, \mnras, 280, 67

\bibitem[{{H.~E.~S.~S. Collaboration} {et~al.}(2021){H.~E.~S.~S.
  Collaboration}, {Abdalla}, {Adam}, {Aharonian}, {Ait Benkhali},
  {Ang{\"u}ner}, {Arcaro}, {Armand}, {Armstrong}, {Ashkar}, {Backes},
  {Baghmanyan}, {Barbosa Martins}, {Barnacka}, {Barnard}, {Becherini}, {Berge},
  {Bernl{\"o}hr}, {Bi}, {B{\"o}ttcher}, {Boisson}, {Bolmont}, {Bonnefoy}, {de
  Bony de Lavergne}, {Bregeon}, {Breuhaus}, {Brun}, {Brun}, {Bryan},
  {B{\"u}chele}, {Bulik}, {Bylund}, {Caroff}, {Carosi}, {Casanova}, {Chand},
  {Chandra}, {Chen}, {Cotter}, {Cury{\l}o}, {Damascene Mbarubucyeye}, {Davids},
  {Davies}, {Deil}, {Devin}, {Dewilt}, {Dirson}, {Djannati-Ata{\"\i}},
  {Dmytriiev}, {Donath}, {Doroshenko}, {Dyks}, {Egberts}, {Eichhorn},
  {Einecke}, {Emery}, {Ernenwein}, {Feijen}, {Fegan}, {Fiasson}, {Fichet de
  Clairfontaine}, {Filipovic}, {Fontaine}, {Funk}, {F{\"u}{\ss}ling}, {Gabici},
  {Gallant}, {Giavitto}, {Giunti}, {Glawion}, {Glicenstein}, {Gottschall},
  {Grondin}, {Hahn}, {Haupt}, {Hermann}, {Hinton}, {Hofmann}, {Hoischen},
  {Holch}, {Holler}, {H{\"o}rbe}, {Horns}, {Huber}, {Jamrozy}, {Jankowsky},
  {Jankowsky}, {Jardin-Blicq}, {Joshi}, {Jung-Richardt}, {Kastendieck},
  {Katarzy{\'n}ski}, {Katz}, {Khangulyan}, {Kh{\'e}lifi}, {Klepser},
  {Klu{\'z}niak}, {Komin}, {Konno}, {Kosack}, {Kostunin}, {Kreter}, {Lamanna},
  {Lemi{\`e}re}, {Lemoine-Goumard}, {Lenain}, {Levy}, {Lohse}, {Lypova},
  {Mackey}, {Majumdar}, {Malyshev}, {Malyshev}, {Marandon}, {Marchegiani},
  {Marcowith}, {Mares}, {Mart{\'\i}-Devesa}, {Marx}, {Maurin}, {Meintjes},
  {Meyer}, {Mitchell}, {Moderski}, {Mohamed}, {Mohrmann}, {Montanari}, {Moore},
  {Morris}, {Moulin}, {Muller}, {Murach}, {Nakashima}, {Nayerhoda}, {de
  Naurois}, {Ndiyavala}, {Niederwanger}, {Niemiec}, {Oakes}, {O'Brien},
  {Odaka}, {Ohm}, {Olivera-Nieto}, {de Ona Wilhelmi}, {Ostrowski}, {Panter},
  {Panny}, {Parsons}, {Peron}, {Peyaud}, {Piel}, {Pita}, {Poireau}, {Priyana
  Noel}, {Prokhorov}, {Prokoph}, {P{\"u}hlhofer}, {Punch}, {Quirrenbach},
  {Raab}, {Rauth}, {Reichherzer}, {Reimer}, {Reimer}, {Remy}, {Renaud},
  {Rieger}, {Rinchiuso}, {Romoli}, {Rowell}, {Rudak}, {Ruiz-Velasco},
  {Sahakian}, {Sailer}, {Sanchez}, {Santangelo}, {Sasaki}, {Scalici},
  {Sch{\"u}ssler}, {Schutte}, {Schwanke}, {Schwemmer}, {Seglar-Arroyo},
  {Senniappan}, {Seyffert}, {Shafi}, {Shiningayamwe}, {Simoni}, {Sinha}, {Sol},
  {Specovius}, {Spencer}, {Spir-Jacob}, {Stawarz}, {Sun}, {Steenkamp},
  {Stegmann}, {Steinmassl}, {Steppa}, {Takahashi}, {Tavernier}, {Taylor},
  {Terrier}, {Tiziani}, {Tluczykont}, {Tomankova}, {Trichard}, {Tsirou},
  {Tuffs}, {Uchiyama}, {van der Walt}, {van Eldik}, {van Rensburg}, {van
  Soelen}, {Vasileiadis}, {Veh}, {Venter}, {Vincent}, {Vink}, {V{\"o}lk},
  {Vuillaume}, {Wadiasingh}, {Wagner}, {Watson}, {Werner}, {White},
  {Wierzcholska}, {Wong}, {Yusafzai}, {Zacharias}, {Zanin}, {Zargaryan},
  {Zdziarski}, {Zech}, {Zhu}, {Zorn}, {Zouari}, {{\.Z}ywucka}, {MAGIC
  Collaboration}, {Acciari}, {Ansoldi}, {Antonelli}, {Arbet Engels}, {Asano},
  {Baack}, {Babi{\'c}}, {Baquero}, {Barres de Almeida}, {Barrio}, {Becerra
  Gonz{\'a}lez}, {Bednarek}, {Bellizzi}, {Bernardini}, {Berti}, {Besenrieder},
  {Bhattacharyya}, {Bigongiari}, {Biland}, {Blanch}, {Bonnoli},
  {Bo{\v{s}}njak}, {Busetto}, {Carosi}, {Ceribella}, {Cerruti}, {Chai},
  {Chilingarian}, {Cikota}, {Colak}, {Colin}, {Colombo}, {Contreras},
  {Cortina}, {Covino}, {D'Amico}, {D'Elia}, {da Vela}, {Dazzi}, {de Angelis},
  {de Lotto}, {Delfino}, {Delgado}, {Depaoli}, {di Pierro}, {di Venere}, {Do
  Souto Espi{\~n}eira}, {Dominis Prester}, {Donini}, {Dorner}, {Doro},
  {Elsaesser}, {Fallah Ramazani}, {Fattorini}, {Ferrara}, {Foffano}, {Fonseca},
  {Font}, {Fruck}, {Fukami}, {Garc{\'\i}a L{\'o}pez}, {Garczarczyk},
  {Gasparyan}, {Gaug}, {Giglietto}, {Giordano}, {Gliwny}, {Godinovi{\'c}},
  {Green}, {Hadasch}, {Hahn}, {Heckmann}, {Herrera}, {Hoang}, {Hrupec},
  {H{\"u}tten}, {Inada}, {Inoue}, {Ishio}, {Iwamura}, {Jouvin}, {Kajiwara},
  {Karjalainen}, {Kerszberg}, {Kobayashi}, {Kubo}, {Kushida}, {Lamastra},
  {Lelas}, {Leone}, {Lindfors}, {Lombardi}, {Longo}, {L{\'o}pez},
  {L{\'o}pez-Coto}, {L{\'o}pez-Oramas}, {Loporchio}, {Machado de Oliveira
  Fraga}, {Maggio}, {Majumdar}, {Makariev}, {Mallamaci}, {Maneva}, {Manganaro},
  {Mannheim}, {Maraschi}, {Mariotti}, {Mart{\'\i}nez}, {Mazin}, {Mender},
  {Mi{\'c}anovi{\'c}}, {Miceli}, {Miener}, {Minev}, {Miranda}, {Mirzoyan},
  {Molina}, {Moralejo}, {Morcuende}, {Moreno}, {Moretti}, {Munar-Adrover},
  {Neustroev}, {Nigro}, {Nilsson}, {Ninci}, {Nishijima}, {Noda}, {Nozaki},
  {Ohtani}, {Oka}, {Otero-Santos}, {Palatiello}, {Paneque}, {Paoletti},
  {Paredes}, {Pavleti{\'c}}, {Pe{\~n}il}, {Perennes}, {Persic}, {Prada Moroni},
  {Prandini}, {Priyadarshi}, {Puljak}, {Rhode}, {Rib{\'o}}, {Rico}, {Righi},
  {Rugliancich}, {Saha}, {Sahakyan}, {Saito}, {Sakurai}, {Satalecka},
  {Schleicher}, {Schmidt}, {Schweizer}, {Sitarek}, {{\v{S}}nidari{\'c}},
  {Sobczynska}, {Spolon}, {Stamerra}, {Strom}, {Strzys}, {Suda}, {Suri{\'c}},
  {Takahashi}, {Tavecchio}, {Temnikov}, {Terzi{\'c}}, {Teshima},
  {Torres-Alb{\`a}}, {Tosti}, {Truzzi}, {van Scherpenberg}, {Vanzo}, {Vazquez
  Acosta}, {Ventura}, {Verguilov}, {Vigorito}, {Vitale}, {Vovk}, {Will},
  {Zari{\'c}}, {Jorstad}, {Marscher}, {Boccardi}, {Casadio}, {Hodgson}, {Kim},
  {Krichbaum}, {L{\"a}hteenm{\"a}ki}, {Tornikoski}, {Traianou}, \&
  {Weaver}}]{Hess21}
{H.~E.~S.~S. Collaboration}, {Abdalla}, H., {Adam}, R., {et~al.} 2021, \aap,
  648, A23

\bibitem[{Harris {et~al.}(2020)Harris, Millman, van~der Walt, Gommers,
  Virtanen, Cournapeau, Wieser, Taylor, Berg, Smith, Kern, Picus, Hoyer, van
  Kerkwijk, Brett, Haldane, del R{'{\i}}o, Wiebe, Peterson,
  G{'{e}}rard-Marchant, Sheppard, Reddy, Weckesser, Abbasi, Gohlke, \&
  Oliphant}]{Harris20}
Harris, C.~R., Millman, K.~J., van~der Walt, S.~J., {et~al.} 2020, Nature, 585,
  357

\bibitem[{{Hewett} \& {Wild}(2010)}]{Hewett10}
{Hewett}, P.~C. \& {Wild}, V. 2010, \mnras, 405, 2302

\bibitem[{{Hovatta} \& {Lindfors}(2019)}]{Hovatta19b}
{Hovatta}, T. \& {Lindfors}, E. 2019, \nar, 87, 101541

\bibitem[{Hunter(2007)}]{Hunter07}
Hunter, J.~D. 2007, Computing in Science \& Engineering, 9, 90

\bibitem[{{Jorstad} \& {Marscher}(2004)}]{Jorstad04b}
{Jorstad}, S.~G. \& {Marscher}, A.~P. 2004, \apj, 614, 615

\bibitem[{{Jorstad} \& {Marscher}(2006)}]{Jorstad06}
{Jorstad}, S.~G. \& {Marscher}, A.~P. 2006, Astronomische Nachrichten, 327, 227

\bibitem[{{Jorstad} {et~al.}(2001){Jorstad}, {Marscher}, {Mattox}, {Aller},
  {Aller}, {Wehrle}, \& {Bloom}}]{Jorstad01}
{Jorstad}, S.~G., {Marscher}, A.~P., {Mattox}, J.~R., {et~al.} 2001, \apj, 556,
  738

\bibitem[{{Jorstad} {et~al.}(2017){Jorstad}, {Marscher}, {Morozova},
  {Troitsky}, {Agudo}, {Casadio}, {Foord}, {G{\'o}mez}, {MacDonald}, {Molina},
  {L{\"a}hteenm{\"a}ki}, {Tammi}, \& {Tornikoski}}]{Jorstad17}
{Jorstad}, S.~G., {Marscher}, A.~P., {Morozova}, D.~A., {et~al.} 2017, \apj,
  846, 98

\bibitem[{{K\"onigl}(1981)}]{Konigl81}
{K\"onigl}, A. 1981, \apj, 243, 700

\bibitem[{{Kouch} {et~al.}(2024){Kouch}, {Liodakis}, {Fenu}, {Zhang}, {Boula},
  {Middei}, {Di Gesu}, {Paraschos}, {Agudo}, {Jorstad}, {Lindfors}, {Marscher},
  {Krawczynski}, {Negro}, {Hu}, {Kim}, {Cavazzuti}, {Errando}, {Blinov},
  {Gourni}, {Kiehlmann}, {Kourtidis}, {Mandarakas}, {Triantafyllou},
  {Vervelaki}, {Borman}, {Kopatskaya}, {Larionova}, {Morozova}, {Savchenko},
  {Vasilyev}, {Troitskiy}, {Grishina}, {Zhovtan}, {Jos{\'e} Aceituno},
  {Bonnoli}, {Casanova}, {Escudero}, {Ag{\'\i}s-Gonz{\'a}lez}, {Husillos},
  {Otero-Santos}, {Piirola}, {Sota}, {Myserlis}, {Gurwell}, {Keating}, {Rao},
  {Angelakis}, {Kraus}, {Antonelli}, {Bachetti}, {Baldini}, {Baumgartner},
  {Bellazzini}, {Bianchi}, {Bongiorno}, {Bonino}, {Brez}, {Bucciantini},
  {Capitanio}, {Castellano}, {Chen}, {Ciprini}, {Costa}, {De Rosa}, {Del
  Monte}, {Di Lalla}, {Di Marco}, {Donnarumma}, {Doroshenko}, {Dov{\v{c}}iak},
  {Ehlert}, {Enoto}, {Evangelista}, {Fabiani}, {Ferrazzoli}, {Garcia}, {Gunji},
  {Hayashida}, {Heyl}, {Iwakiri}, {Kaaret}, {Karas}, {Kislat}, {Kitaguchi},
  {Kolodziejczak}, {La Monaca}, {Latronico}, {Maldera}, {Manfreda}, {Marin},
  {Marinucci}, {Marshall}, {Massaro}, {Matt}, {Mitsuishi}, {Mizuno}, {Muleri},
  {Ng}, {O'Dell}, {Omodei}, {Oppedisano}, {Papitto}, {Pavlov}, {Peirson},
  {Perri}, {Pesce-Rollins}, {Petrucci}, {Pilia}, {Possenti}, {Poutanen},
  {Puccetti}, {Ramsey}, {Rankin}, {Ratheesh}, {Roberts}, {Sgr{\`o}}, {Slane},
  {Soffitta}, {Spandre}, {Swartz}, {Tamagawa}, {Tavecchio}, {Taverna},
  {Tawara}, {Tennant}, {Thomas}, {Tombesi}, {Trois}, {Tsygankov}, {Turolla},
  {Romani}, {Vink}, {Weisskopf}, {Wu}, {Xie}, \& {Zane}}]{Kouch24}
{Kouch}, P.~M., {Liodakis}, I., {Fenu}, F., {et~al.} 2024, arXiv e-prints,
  arXiv:2411.16868

\bibitem[{{Laing}(1996)}]{Laing96}
{Laing}, R.~A. 1996, in Astronomical Society of the Pacific Conference Series,
  Vol. 100, Energy Transport in Radio Galaxies and Quasars, ed. P.~E. {Hardee},
  A.~H. {Bridle}, \& J.~A. {Zensus}, 241

\bibitem[{{Larionov} {et~al.}(2013){Larionov}, {Jorstad}, {Marscher},
  {Morozova}, {Blinov}, {Hagen-Thorn}, {Konstantinova}, {Kopatskaya},
  {Larionova}, {Larionova}, \& {Troitsky}}]{Larionov13}
{Larionov}, V.~M., {Jorstad}, S.~G., {Marscher}, A.~P., {et~al.} 2013, \apj,
  768, 40

\bibitem[{{Le{\'o}n-Tavares} {et~al.}(2011){Le{\'o}n-Tavares}, {Valtaoja},
  {Tornikoski}, {L{\"a}hteenm{\"a}ki}, \& {Nieppola}}]{LeonTavares11}
{Le{\'o}n-Tavares}, J., {Valtaoja}, E., {Tornikoski}, M.,
  {L{\"a}hteenm{\"a}ki}, A., \& {Nieppola}, E. 2011, \aap, 532, A146

\bibitem[{{Liodakis} {et~al.}(2020){Liodakis}, {Blinov}, {Jorstad}, {Arkharov},
  {Di Paola}, {Efimova}, {Grishina}, {Kiehlmann}, {Kopatskaya}, {Larionov},
  {Larionova}, {Larionova}, {Marscher}, {Morozova}, {Nikiforova}, {Pavlidou},
  {Traianou}, {Troitskaya}, {Troitsky}, {Uemura}, \& {Weaver}}]{Liodakis20}
{Liodakis}, I., {Blinov}, D., {Jorstad}, S.~G., {et~al.} 2020, \apj, 902, 61

\bibitem[{{Liodakis} {et~al.}(2022){Liodakis}, {Marscher}, {Agudo},
  {Berdyugin}, {Bernardos}, {Bonnoli}, {Borman}, {Casadio}, {Casanova},
  {Cavazzuti}, {Rodriguez Cavero}, {Di Gesu}, {Di Lalla}, {Donnarumma},
  {Ehlert}, {Errando}, {Escudero}, {Garc{\'\i}a-Comas},
  {Ag{\'\i}s-Gonz{\'a}lez}, {Husillos}, {Jormanainen}, {Jorstad}, {Kagitani},
  {Kopatskaya}, {Kravtsov}, {Krawczynski}, {Lindfors}, {Larionova}, {Madejski},
  {Marin}, {Marchini}, {Marshall}, {Morozova}, {Massaro}, {Masiero}, {Mawet},
  {Middei}, {Millar-Blanchaer}, {Myserlis}, {Negro}, {Nilsson}, {O'Dell},
  {Omodei}, {Pacciani}, {Paggi}, {Panopoulou}, {Peirson}, {Perri}, {Petrucci},
  {Poutanen}, {Puccetti}, {Romani}, {Sakanoi}, {Savchenko}, {Sota},
  {Tavecchio}, {Tinyanont}, {Vasilyev}, {Weaver}, {Zhovtan}, {Antonelli},
  {Bachetti}, {Baldini}, {Baumgartner}, {Bellazzini}, {Bianchi}, {Bongiorno},
  {Bonino}, {Brez}, {Bucciantini}, {Capitanio}, {Castellano}, {Ciprini},
  {Costa}, {De Rosa}, {Del Monte}, {Di Marco}, {Doroshenko}, {Dov{\v{c}}iak},
  {Enoto}, {Evangelista}, {Fabiani}, {Ferrazzoli}, {Garcia}, {Gunji},
  {Hayashida}, {Heyl}, {Iwakiri}, {Karas}, {Kitaguchi}, {Kolodziejczak}, {La
  Monaca}, {Latronico}, {Maldera}, {Manfreda}, {Marinucci}, {Matt},
  {Mitsuishi}, {Mizuno}, {Muleri}, {Ng}, {Oppedisano}, {Papitto}, {Pavlov},
  {Pesce-Rollins}, {Pilia}, {Possenti}, {Ramsey}, {Rankin}, {Ratheesh},
  {Sgr{\'o}}, {Slane}, {Soffitta}, {Spandre}, {Tamagawa}, {Taverna}, {Tawara},
  {Tennant}, {Thomas}, {Tombesi}, {Trois}, {Tsygankov}, {Turolla}, {Vink},
  {Weisskopf}, {Wu}, {Xie}, \& {Zane}}]{Liodakis22}
{Liodakis}, I., {Marscher}, A.~P., {Agudo}, I., {et~al.} 2022, \nat, 611, 677

\bibitem[{{MacDonald} {et~al.}(2017){MacDonald}, {Jorstad}, \&
  {Marscher}}]{MacDonald17}
{MacDonald}, N.~R., {Jorstad}, S.~G., \& {Marscher}, A.~P. 2017, \apj, 850, 87

\bibitem[{{MacDonald} {et~al.}(2015){MacDonald}, {Marscher}, {Jorstad}, \&
  {Joshi}}]{MacDonald15}
{MacDonald}, N.~R., {Marscher}, A.~P., {Jorstad}, S.~G., \& {Joshi}, M. 2015,
  \apj, 804, 111

\bibitem[{{Marchesini} {et~al.}(2016){Marchesini}, {Andruchow}, {Cellone},
  {Combi}, {Zibecchi}, {Mart{\'\i}}, {Romero}, {Mu{\~n}oz-Arjonilla},
  {Luque-Escamilla}, \& {S{\'a}nchez-Sutil}}]{Marchesini16}
{Marchesini}, E.~J., {Andruchow}, I., {Cellone}, S.~A., {et~al.} 2016, \aap,
  591, A21

\bibitem[{{Marscher} \& {Broderick}(1983)}]{Marscher83b}
{Marscher}, A.~P. \& {Broderick}, J.~J. 1983, \aj, 88, 759

\bibitem[{{Marscher} \& {Gear}(1985)}]{Marscher85}
{Marscher}, A.~P. \& {Gear}, W.~K. 1985, \apj, 298, 114

\bibitem[{{Marscher} {et~al.}(2008){Marscher}, {Jorstad}, {D'Arcangelo},
  {Smith}, {Williams}, {Larionov}, {Oh}, {Olmstead}, {Aller}, {Aller},
  {McHardy}, {L{\"a}hteenm{\"a}ki}, {Tornikoski}, {Valtaoja}, {Hagen-Thorn},
  {Kopatskaya}, {Gear}, {Tosti}, {Kurtanidze}, {Nikolashvili}, {Sigua},
  {Miller}, \& {Ryle}}]{Marscher08}
{Marscher}, A.~P., {Jorstad}, S.~G., {D'Arcangelo}, F.~D., {et~al.} 2008, \nat,
  452, 966

\bibitem[{{Marscher} {et~al.}(2002){Marscher}, {Jorstad}, {Mattox}, \&
  {Wehrle}}]{Marscher02}
{Marscher}, A.~P., {Jorstad}, S.~G., {Mattox}, J.~R., \& {Wehrle}, A.~E. 2002,
  \apj, 577, 85

\bibitem[{{McCall} {et~al.}(2024){McCall}, {Jermak}, {Steele}, {Kobayashi},
  {Knapen}, \& {S{\'a}nchez-Alarc{\'o}n}}]{McCall24}
{McCall}, C., {Jermak}, H.~E., {Steele}, I.~A., {et~al.} 2024, \mnras, 528,
  4702

\bibitem[{{Nalewajko}(2017)}]{Nalewajko17}
{Nalewajko}, K. 2017, Galaxies, 5, 64

\bibitem[{{Paraschos} {et~al.}(2024{\natexlab{a}}){Paraschos}, {Debbrecht},
  {Kramer}, {Traianou}, {Liodakis}, {Krichbaum}, {Kim}, {Janssen}, {Nair},
  {Savolainen}, {Ros}, {Bach}, {Hodgson}, {Lisakov}, {MacDonald}, \&
  {Zensus}}]{Paraschos24b}
{Paraschos}, G.~F., {Debbrecht}, L.~C., {Kramer}, J.~A., {et~al.}
  2024{\natexlab{a}}, \aap, 686, L5

\bibitem[{{Paraschos} {et~al.}(2024{\natexlab{b}}){Paraschos}, {Kim},
  {Wielgus}, {R{\"o}der}, {Krichbaum}, {Ros}, {Agudo}, {Myserlis},
  {Moscibrodzka}, {Traianou}, {Zensus}, {Blackburn}, {Chan}, {Issaoun},
  {Janssen}, {Johnson}, {Fish}, {Akiyama}, {Alberdi}, {Alef}, {Algaba},
  {Anantua}, {Asada}, {Azulay}, {Bach}, {Baczko}, {Ball}, {Balokovi{\'c}},
  {Barrett}, {Baub{\"o}ck}, {Benson}, {Bintley}, {Blundell}, {Bouman}, {Bower},
  {Boyce}, {Bremer}, {Brinkerink}, {Brissenden}, {Britzen}, {Broderick},
  {Broguiere}, {Bronzwaer}, {Bustamante}, {Byun}, {Carlstrom}, {Ceccobello},
  {Chael}, {Chang}, {Chatterjee}, {Chatterjee}, {Chen}, {Chen}, {Cheng}, {Cho},
  {Christian}, {Conroy}, {Conway}, {Cordes}, {Crawford}, {Crew}, {Cruz-Osorio},
  {Cui}, {Dahale}, {Davelaar}, {De Laurentis}, {Deane}, {Dempsey}, {Desvignes},
  {Dexter}, {Dhruv}, {Doeleman}, {Dougal}, {Dzib}, {Eatough}, {Emami},
  {Falcke}, {Farah}, {Fomalont}, {Ford}, {Foschi}, {Fraga-Encinas}, {Freeman},
  {Friberg}, {Fromm}, {Fuentes}, {Galison}, {Gammie}, {Garc{\'\i}a}, {Gentaz},
  {Georgiev}, {Goddi}, {Gold}, {G{\'o}mez-Ruiz}, {G{\'o}mez}, {Gu}, {Gurwell},
  {Hada}, {Haggard}, {Haworth}, {Hecht}, {Hesper}, {Heumann}, {Ho}, {Ho},
  {Honma}, {Huang}, {Huang}, {Hughes}, {Ikeda}, {Impellizzeri}, {Inoue},
  {James}, {Jannuzi}, {Jeter}, {Jaing}, {Jim{\'e}nez-Rosales}, {Jorstad},
  {Joshi}, {Jung}, {Karami}, {Karuppusamy}, {Kawashima}, {Keating}, {Kettenis},
  {Kim}, {Kim}, {Kim}, {Kino}, {Koay}, {Kocherlakota}, {Kofuji}, {Koch},
  {Koyama}, {Kramer}, {Kramer}, {Kramer}, {Kuo}, {La Bella}, {Lauer}, {Lee},
  {Lee}, {Leung}, {Levis}, {Li}, {Lico}, {Lindahl}, {Lindqvist}, {Lisakov},
  {Liu}, {Liu}, {Liuzzo}, {Lo}, {Lobanov}, {Loinard}, {Lonsdale}, {Lowitz},
  {Lu}, {MacDonald}, {Mao}, {Marchili}, {Markoff}, {Marrone}, {Marscher},
  {Mart{\'\i}-Vidal}, {Matsushita}, {Matthews}, {Medeiros}, {Menten},
  {Michalik}, {Mizuno}, {Mizuno}, {Moran}, {Moriyama}, {Mulaudzi},
  {M{\"u}ller}, {M{\"u}ller}, {Mus}, {Musoke}, {Nadolski}, {Nagai}, {Nagar},
  {Nakamura}, {Narayanan}, {Natarajan}, {Nathanail}, {Navarro Fuentes},
  {Neilsen}, {Neri}, {Ni}, {Noutsos}, {Nowak}, {Oh}, {Okino}, {Olivares},
  {Ortiz-Le{\'o}n}, {Oyama}, {{\"O}zel}, {Palumbo}, {Park}, {Parsons}, {Patel},
  {Pen}, {Pi{\'e}tu}, {Plambeck}, {PopStefanija}, {Porth}, {P{\"o}tzl},
  {Prather}, {Preciado-L{\'o}pez}, {Psaltis}, {Pu}, {Ramakrishnan}, {Rao},
  {Rawlings}, {Raymond}, {Rezzolla}, {Ricarte}, {Ripperda}, {Roelofs},
  {Rogers}, {Romero-Ca{\~n}izales}, {Roshanineshat}, {Rottmann}, {Roy}, {Ruiz},
  {Ruszczyk}, {Rygl}, {S{\'a}nchez}, {S{\'a}nchez-Arg{\"u}elles},
  {S{\'a}nchez-Portal}, {Sasada}, {Satapathy}, {Savolainen}, {Schloerb},
  {Schonfeld}, {Schuster}, {Shao}, {Shen}, {Small}, {Sohn}, {SooHoo},
  {Sosapanta Salas}, {Souccar}, {Sun}, {Tazaki}, {Tetarenko}, {Tiede},
  {Tilanus}, {Titus}, {Torne}, {Toscano}, {Trent}, {Trippe}, {Turk}, {van
  Bemmel}, {van Langevelde}, {van Rossum}, {Vos}, {Wagner}, {Ward-Thompson},
  {Wardle}, {Washington}, {Weintroub}, {Wharton}, {Wiik}, {Witzel}, {Wondrak},
  {Wong}, {Wu}, {Yadlapalli}, {Yamaguchi}, {Yfantis}, {Yoon}, {Young}, {Young},
  {Younsi}, {Yu}, {Yuan}, {Yuan}, {Zhang}, {Zhao}, \& {Zhao}}]{Paraschos24a}
{Paraschos}, G.~F., {Kim}, J.~Y., {Wielgus}, M., {et~al.} 2024{\natexlab{b}},
  \aap, 682, L3

\bibitem[{{Paraschos} {et~al.}(2022){Paraschos}, {Krichbaum}, {Kim}, {Hodgson},
  {Oh}, {Ros}, {Zensus}, {Marscher}, {Jorstad}, {Gurwell},
  {L{\"a}hteenm{\"a}ki}, {Tornikoski}, {Kiehlmann}, \&
  {Readhead}}]{Paraschos22}
{Paraschos}, G.~F., {Krichbaum}, T.~P., {Kim}, J.~Y., {et~al.} 2022, \aap, 665,
  A1

\bibitem[{{Paraschos} {et~al.}(2023){Paraschos}, {Mpisketzis}, {Kim}, {Witzel},
  {Krichbaum}, {Zensus}, {Gurwell}, {L{\"a}hteenm{\"a}ki}, {Tornikoski},
  {Kiehlmann}, \& {Readhead}}]{Paraschos23}
{Paraschos}, G.~F., {Mpisketzis}, V., {Kim}, J.~Y., {et~al.} 2023, \aap, 669,
  A32

\bibitem[{{Paraschos} {et~al.}(2024{\natexlab{c}}){Paraschos}, {Wielgus},
  {Benke}, {Mpisketzis}, {R{\"o}sch}, {Dasyra}, {Ros}, {Kadler}, {Ojha},
  {Edwards}, {Hyland}, {Quick}, \& {Weston}}]{Paraschos24c}
{Paraschos}, G.~F., {Wielgus}, M., {Benke}, P., {et~al.} 2024{\natexlab{c}},
  \aap, 687, L6

\bibitem[{{Piner} {et~al.}(2006){Piner}, {Bhattarai}, {Edwards}, \&
  {Jones}}]{Piner06}
{Piner}, B.~G., {Bhattarai}, D., {Edwards}, P.~G., \& {Jones}, D.~L. 2006,
  \apj, 640, 196

\bibitem[{{Price} {et~al.}(1993){Price}, {Gower}, {Hutchings}, {Talon},
  {Duncan}, \& {Ross}}]{Price93}
{Price}, R., {Gower}, A.~C., {Hutchings}, J.~B., {et~al.} 1993, \apjs, 86, 365

\bibitem[{{Raiteri} {et~al.}(1998){Raiteri}, {Ghisellini}, {Villata}, {de
  Francesco}, {Lanteri}, {Chiaberge}, {Peila}, \& {Antico}}]{Raiteri98}
{Raiteri}, C.~M., {Ghisellini}, G., {Villata}, M., {et~al.} 1998, \aaps, 127,
  445

\bibitem[{{Rani} {et~al.}(2018){Rani}, {Jorstad}, {Marscher}, {Agudo},
  {Sokolovsky}, {Larionov}, {Smith}, {Mosunova}, {Borman}, {Grishina},
  {Kopatskaya}, {Mokrushina}, {Morozova}, {Savchenko}, {Troitskaya},
  {Troitsky}, {Thum}, {Molina}, \& {Casadio}}]{Rani18}
{Rani}, B., {Jorstad}, S.~G., {Marscher}, A.~P., {et~al.} 2018, \apj, 858, 80

\bibitem[{{Roelofs} {et~al.}(2023){Roelofs}, {Johnson}, {Chael}, {Janssen},
  {Wielgus}, {Broderick}, {Akiyama}, {Alberdi}, {Alef}, {Algaba}, {Anantua},
  {Asada}, {Azulay}, {Bach}, {Baczko}, {Ball}, {Balokovi{\'c}}, {Barrett},
  {Baub{\"o}ck}, {Benson}, {Bintley}, {Blackburn}, {Blundell}, {Bouman},
  {Bower}, {Boyce}, {Bremer}, {Brinkerink}, {Brissenden}, {Britzen},
  {Broguiere}, {Bronzwaer}, {Bustamante}, {Byun}, {Carlstrom}, {Ceccobello},
  {Chan}, {Chang}, {Chatterjee}, {Chatterjee}, {Chen}, {Chen}, {Cheng}, {Cho},
  {Christian}, {Conroy}, {Conway}, {Cordes}, {Crawford}, {Crew}, {Cruz-Osorio},
  {Cui}, {Dahale}, {Davelaar}, {De Laurentis}, {Deane}, {Dempsey}, {Desvignes},
  {Dexter}, {Dhruv}, {Doeleman}, {Dougal}, {Dzib}, {Eatough}, {Emami},
  {Falcke}, {Farah}, {Fish}, {Fomalont}, {Ford}, {Foschi}, {Fraga-Encinas},
  {Freeman}, {Friberg}, {Fromm}, {Fuentes}, {Galison}, {Gammie}, {Garc{\'\i}a},
  {Gentaz}, {Georgiev}, {Goddi}, {Gold}, {G{\'o}mez-Ruiz}, {G{\'o}mez}, {Gu},
  {Gurwell}, {Hada}, {Haggard}, {Haworth}, {Hecht}, {Hesper}, {Heumann}, {Ho},
  {Ho}, {Honma}, {Huang}, {Huang}, {Hughes}, {Ikeda}, {Impellizzeri}, {Inoue},
  {Issaoun}, {James}, {Jannuzi}, {Jeter}, {Jiang}, {Jim{\'e}nez-Rosales},
  {Jorstad}, {Joshi}, {Jung}, {Karami}, {Karuppusamy}, {Kawashima}, {Keating},
  {Kettenis}, {Kim}, {Kim}, {Kim}, {Kim}, {Kino}, {Koay}, {Kocherlakota},
  {Kofuji}, {Koch}, {Koyama}, {Kramer}, {Kramer}, {Kramer}, {Krichbaum}, {Kuo},
  {La Bella}, {Lauer}, {Lee}, {Lee}, {Leung}, {Levis}, {Li}, {Lico}, {Lindahl},
  {Lindqvist}, {Lisakov}, {Liu}, {Liu}, {Liuzzo}, {Lo}, {Lobanov}, {Loinard},
  {Lonsdale}, {Lowitz}, {Lu}, {MacDonald}, {Mao}, {Marchili}, {Markoff},
  {Marrone}, {Marscher}, {Mart{\'\i}-Vidal}, {Matsushita}, {Matthews},
  {Medeiros}, {Menten}, {Michalik}, {Mizuno}, {Mizuno}, {Moran}, {Moriyama},
  {Moscibrodzka}, {Mulaudzi}, {M{\"u}ller}, {M{\"u}ller}, {Mus}, {Musoke},
  {Myserlis}, {Nadolski}, {Nagai}, {Nagar}, {Nakamura}, {Narayanan},
  {Natarajan}, {Nathanail}, {Fuentes}, {Neilsen}, {Neri}, {Ni}, {Noutsos},
  {Nowak}, {Oh}, {Okino}, {Olivares}, {Ortiz-Le{\'o}n}, {Oyama}, {{\"O}zel},
  {Palumbo}, {Paraschos}, {Park}, {Parsons}, {Patel}, {Pen}, {Pesce},
  {Pi{\'e}tu}, {Plambeck}, {PopStefanija}, {Porth}, {P{\"o}tzl}, {Prather},
  {Preciado-L{\'o}pez}, {Psaltis}, {Pu}, {Ramakrishnan}, {Rao}, {Rawlings},
  {Raymond}, {Rezzolla}, {Ricarte}, {Ripperda}, {Rogers},
  {Romero-Ca{\~n}izales}, {Ros}, {Roshanineshat}, {Rottmann}, {Roy}, {Ruiz},
  {Ruszczyk}, {Rygl}, {S{\'a}nchez}, {S{\'a}nchez-Arg{\"u}elles},
  {S{\'a}nchez-Portal}, {Sasada}, {Satapathy}, {Savolainen}, {Schloerb},
  {Schonfeld}, {Schuster}, {Shao}, {Shen}, {Small}, {Sohn}, {SooHoo},
  {Sosapanta Salas}, {Souccar}, {Sun}, {Tazaki}, {Tetarenko}, {Tiede},
  {Tilanus}, {Titus}, {Torne}, {Toscano}, {Traianou}, {Trent}, {Trippe},
  {Turk}, {van Bemmel}, {van Langevelde}, {van Rossum}, {Vos}, {Wagner},
  {Ward-Thompson}, {Wardle}, {Washington}, {Weintroub}, {Wharton}, {Wiik},
  {Witzel}, {Wondrak}, {Wong}, {Wu}, {Yadlapalli}, {Yamaguchi}, {Yfantis},
  {Yoon}, {Young}, {Young}, {Younsi}, {Yu}, {Yuan}, {Yuan}, {Zensus}, {Zhang},
  {Zhao}, \& {Zhao}}]{Roelofs23}
{Roelofs}, F., {Johnson}, M.~D., {Chael}, A., {et~al.} 2023, \apjl, 957, L21

\bibitem[{{Saito} {et~al.}(2015){Saito}, {Stawarz}, {Tanaka}, {Takahashi},
  {Sikora}, \& {Moderski}}]{Saito15}
{Saito}, S., {Stawarz}, {\L}., {Tanaka}, Y.~T., {et~al.} 2015, \apj, 809, 171

\bibitem[{{Sol} {et~al.}(1989){Sol}, {Pelletier}, \& {Asseo}}]{Sol89}
{Sol}, H., {Pelletier}, G., \& {Asseo}, E. 1989, \mnras, 237, 411

\bibitem[{{Villata} {et~al.}(1997){Villata}, {Raiteri}, {Ghisellini}, {de
  Francesco}, {Bosio}, {Latini}, {Bucciarelli}, {Chiaberge}, {Chiumiento},
  {Cora}, {Lanteri}, {Lattanzi}, {Massone}, {Peila}, {Racioppi}, {Smart}, \&
  {Scaltriti}}]{Villata97}
{Villata}, M., {Raiteri}, C.~M., {Ghisellini}, G., {et~al.} 1997, \aaps, 121,
  119

\bibitem[{Virtanen {et~al.}(2020)Virtanen, Gommers, Oliphant, Haberland, Reddy,
  Cournapeau, Burovski, Peterson, Weckesser, Bright, {van der Walt}, Brett,
  Wilson, Millman, Mayorov, Nelson, Jones, Kern, Larson, Carey, Polat, Feng,
  Moore, {VanderPlas}, Laxalde, Perktold, Cimrman, Henriksen, Quintero, Harris,
  Archibald, Ribeiro, Pedregosa, {van Mulbregt}, \& {SciPy 1.0
  Contributors}}]{2020SciPy-NMeth}
Virtanen, P., Gommers, R., Oliphant, T.~E., {et~al.} 2020, Nature Methods, 17,
  261

\bibitem[{{Zhang} {et~al.}(2024){Zhang}, {Xiong}, {Gao}, {Yang}, {Lu}, {Na}, \&
  {Qin}}]{Zhang24}
{Zhang}, X., {Xiong}, D.-r., {Gao}, Q.-g., {et~al.} 2024, \mnras, 529, 3699

\end{thebibliography}

\end{document}